\documentclass[12pt]{article}

\usepackage{graphics,amssymb,epsfig,float}
\usepackage[usenames,dvips]{color}
\usepackage{graphicx}
\usepackage{epsfig}
\usepackage{rotating}
\usepackage{dcolumn}
\usepackage{bm}
\usepackage{cite}
\usepackage{amsmath}

\def\lsim{\;\raise0.3ex\hbox{$<$\kern-0.75em\raise-1.1ex\hbox{$\sim$}}\;}
\def\gsim{\;\raise0.3ex\hbox{$>$\kern-0.75em\raise-1.1ex\hbox{$\sim$}}\;}
\def\beq{\begin{equation}}   \def\eeq{\end{equation}}
\def\ba{\begin{array}}       \def\ea{\end{array}}
\def\bea{\begin{eqnarray}}   \def\eea{\end{eqnarray}}

\def\nl{\newline}
\def\k{\kappa}
\def\l{\lambda}

\begin{document}

\begin{titlepage}
\begin{flushright}
LPT Orsay 12-54
\end{flushright}

\vspace{1cm}
\begin{center}
{\Large\bf A 130 GeV photon line from dark matter annihilation}

\vspace{5mm}
{\Large\bf in the NMSSM} 
\vspace{2cm}

{\bf{Debottam Das, Ulrich Ellwanger and Pantelis Mitropoulos}}
\vspace{1cm}\\
\it  Laboratoire de Physique Th\'eorique, UMR 8627, CNRS and
Universit\'e de Paris--Sud,\\
\it B\^at. 210, 91405 Orsay, France
\end{center}

\vspace{1cm}
\begin{abstract}
In the Next-to-Minimal Supersymmetric Standard Model, neutralino dark
matter can annihilate into a pair of photons through the exchange of a
CP-odd Higgs boson in the s-channel. The CP-odd Higgs boson couples to
two photons through a loop of dominantly higgsino-like charginos. We
show that the parameter space of the NMSSM can accommodate simultaneously
i)~neutralino-like dark matter of a mass of about 130~GeV giving rise to
a 130~GeV photon line; ii)~an annihilation cross section of or larger
than $10^{-27}$~cm$^3$~s$^{-1}$; iii)~a relic density in agreement with
WMAP constraints; iv)~a direct detection cross section compatible with
bounds from XENON100, and v)~a Standard Model like Higgs mass of about
125~GeV. However, the CP-odd Higgs mass has to lie accidentally close
to 260~GeV.

\end{abstract}

\end{titlepage}

\section{Introduction}

The fact that annihilation of dark matter can give rise to photons
with a well-defined energy
\cite{Bergstrom:1988fp,Bergstrom:1997fh,Bern:1997ng,Ullio:1997ke,
Bergstrom:1997fj,
Bergstrom:2004nr,Boudjema:2005hb,Birkedal:2006fz,Ferrer:2006hy,
Perelstein:2006bq,Gustafsson:2007pc,Bertone:2009cb,Dudas:2009uq,
Mambrini:2009ad,Jackson:2009kg,Goodman:2010qn,Bertone:2010fn,
Profumo:2010kp,Bringmann:2011ye,Chalons:2011ia,Chalons:2012hf,
Bergstrom:2012fi}
has motivated searches for such gamma lines originating from the center
of our galaxy by the EGRET~\cite{Pullen:2006sy},
H.E.S.S.~\cite{Aharonian:2006wh,Abramowski:2011hc} and
Fermi~LAT~\cite{Abdo:2010nc,Vertongen:2011mu,Ackermann:2012qk}
experiments. 

Recent analyses of the publicly available data from Fermi~LAT
\cite{Bringmann:2012vr,Weniger:2012tx,Tempel:2012ey,{1206.1616}} have
discovered hints for a gamma line at $E_\gamma \sim 130$~GeV in the form
of an excess of about $3-4$~standard deviations, assuming that the
background flux can be approximated by a single power law. An
interpretation of this excess as dark matter pair annihilation into a
pair of photons would require a partial annihilation cross section of
about $10^{-27}$~cm$^3$~s$^{-1}$ \cite{Weniger:2012tx,Tempel:2012ey}.

However, more general parametrizations of the background flux
\cite{Profumo:2012tr,Boyarsky:2012ca} reduce the significance of the
excess, making it compatible with a diffuse background possibly of
instrumental or astrophysical origin. The Fermi~LAT collaboration
preferred to interpret the same data only in terms of upper bounds on
the partial annihilation cross section \cite{Ackermann:2012qk}.

Still, future additional data could confirm the present hints for a
possible excess; hence it is of interest to study whether it could be
explained within concrete models for dark matter which are compatible
with bounds on its relic density from WMAP~\cite{Komatsu:2010fb} and
bounds on its direct detection cross section (in the relevant mass
range) from XENON100~\cite{Aprile:2011hi}.

Following the publication \cite{Weniger:2012tx}, different types of such
models have been proposed: models with an extra U(1)' gauge symmetry
where the dark matter couples only to the extra $Z'$ gauge boson, and
the $Z'$ to photons via a Chern-Simons term \cite{Dudas:2012pb}; models
with an extra singlet and extra charged fields allowing for an enhanced
branching ratio of the Standard Model-like Higgs boson into two photons
\cite{Cline:2012nw}; extensions of the Minimal Supersymmetric Standard
Model (MSSM) by right-handed neutrinos and extra Higgs doublets (with a
right-handed sneutrino as dark matter) \cite{Choi:2012ap}; decaying dark
matter in the MSSM extended by additional fields and
couplings~\cite{Kyae:2012vi}; extensions of the Standard Model or the
MSSM by singlets with Peccei-Quinn symmetry where dark matter can
annihilate into a photon pair via an axion in the s-channel
\cite{Lee:2012bq}; a string/M-theory motivated version of the MSSM (with
an unconventional spectrum involving very heavy scalar superpartners)
where wino-like dark matter can annihilate into a photon plus a $Z$
boson \cite{Acharya:2012dz}. (In the MSSM with bino- or higgsino-like
dark matter, the annihilation cross section into one or two photons
would be too small \cite{Bergstrom:1997fh, Bern:1997ng,
Ullio:1997ke,Bergstrom:1997fj, Boudjema:2005hb}.)

In the present paper we consider the simplest supersymmetric extension
of the Standard Model with a scale invariant superpotential, the
Next-to-Minimal Supersymmetric Standard Model (NMSSM)
\cite{Maniatis:2009re,Ellwanger:2009dp}.  In the NMSSM, the
Higgs/higgsino mass term $\mu H_u H_d$ in the superpotential of the MSSM
is replaced by a coupling $\lambda S H_u H_d$ to a gauge singlet
superfield~$S$. Compared to the MSSM, the Higgs sector of the NMSSM
includes additional neutral CP-even and CP-odd states. It is known that
a large dark matter (neutralino) annihilation cross section into two
photons can arise from neutralino annihilation through the
NMSSM-specific CP-odd Higgs boson in the s-channel
\cite{Ferrer:2006hy,Chalons:2011ia, Chalons:2012hf,Lee:2012bq}.
Moreover, due to the coupling $\lambda S H_u H_d$ the lightest CP-even
Higgs mass is naturally heavier than in the MSSM
\cite{Ellis:1988er,Drees:1988fc, Ellwanger:1993hn,Maniatis:2009re,
Ellwanger:2009dp,Ellwanger:2006rm} which makes it easier to explain a
Higgs mass of 125~GeV \cite{Hall:2011aa,Arvanitaki:2011ck,
Ellwanger:2011aa, Gunion:2012zd, King:2012is,Kang:2012tn,Cao:2012fz,
Vasquez:2012hn,Ellwanger:2012ke, Lodone:2012kp,Jeong:2012ma}.

Hence an obvious question is whether the parameter space of the NMSSM
can describe simultaneously the following phenomena: 1) A
neutralino-like dark matter particle of a mass of about 130~GeV with a
partial annihilation cross section into two photons of about
$10^{-27}$~cm$^3$~s$^{-1}$ and a relic density compatible with WMAP
constraints, 2) a Standard Model-like Higgs boson with a mass of
about 125~GeV, but 3) respecting the upper bound of XENON100 on the dark
matter direct detection cross section.

We find that this is indeed possible, provided the NMSSM-specific
coupling $\lambda$ is relatively large, $\lambda \sim 0.6$. This
coupling plays several r\^oles simultaneously: i) it determines the
coupling of the neutralino-like dark matter to the NMSSM-specific CP-odd
Higgs boson; ii) it determines the coupling of the CP-odd Higgs boson to
charged higgsinos, whose loop leads to a large decay width into two
photons; iii) it helps to increase the mass of the Standard Model-like
CP-even Higgs boson. However, the mass of the NMSSM-specific CP-odd
Higgs boson has to be close to 260~GeV (the s-channel pole) for a
neutralino annihilation cross section into two photons equal to or
larger than $10^{-27}$~cm$^3$~s$^{-1}$ (see \cite{1205.6811}). On the
other hand, no new particles or interactions beyond the NMSSM need to be
introduced if the present hints for a 130~GeV gamma line from dark
matter annihilation in our galaxy are confirmed.

In the next section we introduce the NMSSM and describe, which
properties of the NMSSM allow to describe simultaneously all the above
phenomena. We indicate a possible region in the parameter space, and
discuss the range of possible direct detection cross sections and
neutralino annihilation cross sections into two photons. Details of the
Higgs and neutralino sector are presented for a typical benchmark point.
Section~3 is devoted to a short summary and conclusions.

\section{A 130 GeV Photon Line in the NMSSM}

The NMSSM differs from the MSSM due to the presence of the gauge singlet
superfield $\widehat S$. In the simplest realisation of the NMSSM, the
Higgs mass term $\mu \widehat H_u \widehat H_d$ in the MSSM
superpotential $W_{\text{MSSM}}$  is replaced by the coupling $\lambda$
of $\widehat S$ to $\widehat H_u$ and $\widehat H_d$, and a
self-coupling $\kappa \widehat S^3$.  Hence, in this version the
superpotential $W_{\text{NMSSM}}$ is scale invariant, and given by:
\beq\label{eq:1}
W_{\text{NMSSM}} = \lambda \widehat S \widehat H_u\cdot \widehat H_d +
\frac{\kappa}{3}  \widehat S^3 +\dots\; ,
\eeq
where the dots denote the Yukawa couplings of $\widehat H_u$ and
$\widehat H_d$ to the quarks and leptons as in the MSSM. 
The NMSSM-specific soft SUSY breaking terms consist of a mass term for
the scalar components of $\widehat S$, and trilinear interactions
associated to the terms in $W_{\text{NMSSM}}$:
\beq\label{eq:2}
 -{\cal L}_{\text{NMSSM}}^{Soft} = m_{S}^2 |S|^2 +\Bigl( \lambda
A_\lambda\, H_u \cdot H_d \,S +  \frac{1}{3} \kappa  A_\kappa\,  S^3
\Bigl)\ +\ \mathrm{h.c.}\;.
\eeq
Subsequently we define the vacuum expectation values (vevs)
\beq\label{eq:3}
\left<H_u\right>=v_u\; ,\quad \left<H_d\right>=v_d\; ,\quad 
\left<S\right>=s\; .
\eeq
In terms of $s$, the first term in
$W_{\text{NMSSM}}$ generates an effective $\mu$-term with
\beq\label{eq:4}
\mu_\mathrm{eff}=\lambda s\; .
\eeq

Using the minimization equations of the potential in order eliminate the
soft SUSY breaking Higgs mass terms, the Higgs sector of the NMSSM is
characterized (at tree level) by the six parameters
\beq\label{eq:5}
\lambda\; ,\quad \kappa\; ,\quad A_\lambda\; ,\quad A_\kappa\; ,\quad
\mu_\mathrm{eff}\; , \quad \tan\beta \equiv \frac{v_u}{v_d}\; .
\eeq

The neutral CP-even Higgs sector contains 3 states $H_i$, which are
mixtures of the CP-even components of the superfields $\widehat H_u$,
$\widehat H_d$ and $\widehat S$. Their masses are described by a $3
\times 3$ mass matrix ${\cal M}^2_{H\,ij}$, where the dominant
contribution to the singlet-like component ${\cal M}^2_{H\,33}$ reads
\cite{Maniatis:2009re,Ellwanger:2009dp}
\beq\label{eq:6}
{\cal M}^2_{H\,33} \sim \kappa s\left(A_\kappa + 4 \kappa s\right)\; .
\eeq

The neutral CP-odd Higgs sector contains 2 physical states $A_i$, whose
masses are described by a $2 \times 2$ mass matrix ${\cal M}^2_{A\,ij}$
where ${\cal M}^2_{A\,11}$ corresponds to the MSSM-like CP-odd Higgs
mass squared. The dominant contributions to the singlet-like component
${\cal M}^2_{A\,22}$ and the singlet-doublet mixing term ${\cal
M}^2_{A\,12}$ are given by
\beq\label{eq:7}
{\cal M}^2_{A\,22} \sim -3\kappa s A_\kappa\; ,\quad
{\cal M}^2_{A\,12} \sim \lambda(A_\lambda-2\kappa s)
\sqrt{v_u^2+v_d^2}\; ,
\eeq
respectively \cite{Maniatis:2009re,Ellwanger:2009dp}.

In the neutralino sector we have 5 states ${\chi}^0_i$, which are
mixtures of the bino $\tilde{B}$, the neutral wino $\tilde{W}^3$, the
neutral higgsinos $\tilde{H_d}^0$, $\tilde{H_u}^0$ from the superfields
$\widehat H_d^0$ and $\widehat H_u^0$, and the singlino from the
superfield $\widehat S$. Their masses are described by a symmetric $5
\times 5$ mass matrix ${\cal M}_{\chi^0\,ij}$ given by
\beq\label{eq:8}
{\cal M}_{\chi^0\,ij} =
\left( \ba{ccccc}
M_1 & 0 & -\frac{g_1 v_d}{\sqrt{2}} & \frac{g_1 v_u}{\sqrt{2}} & 0 \\
& M_2 & \frac{g_2 v_d}{\sqrt{2}} & -\frac{g_2 v_u}{\sqrt{2}} & 0 \\
& & 0 & -\mu_\mathrm{eff} & -\l v_u \\
& & & 0 & -\l v_d \\
& & & & 2 \k s
\ea \right)
\eeq
where $M_1$, $M_2$ are the soft SUSY breaking bino and wino
mass terms.
The lightest eigenstate ${\chi}^0_1$ of the neutralino mass matrix
\eqref{eq:8} is considered as the dark matter particle. In
the above basis, its decomposition is written as
\beq\label{eq:9}
{\chi}^0_1=N_{11} \tilde{B} + N_{12} \tilde{W}^3 + N_{13} \tilde{H_d}^0
 + N_{14} \tilde{H_u}^0 + N_{15} \tilde{S}\; .
 \eeq

\begin{figure}[ht!]
\begin{center}
\includegraphics[scale=0.7,angle=-0]{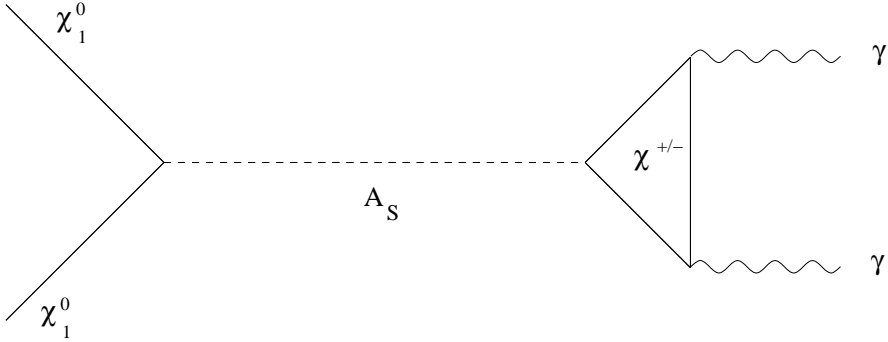}
\end{center}
\caption{The dominant diagram for pair annihilation of ${\chi}^0_1$ into
two photons via a mostly singlet-like CP-odd Higgs $A_S$ in the
NMSSM.}
\label{fig:1}
\end{figure}

The pair annihilation of ${\chi}^0_1$ into two photons can be generated
via a dominantly singlet-like CP-odd Higgs state $A_S$ in the s-channel
\cite{Ferrer:2006hy,Chalons:2011ia, Chalons:2012hf}, see
Fig.~\ref{fig:1}. (Similar diagrams with $\chi^\pm$ replaced by
(s)quarks or (s)leptons are numerically negligible.)

Now we turn to the masses and couplings of the particles and the
corresponding regions in the NMSSM parameter space, which have all
desired phenomenological properties. First we consider the
dominantly singlet-like CP-odd Higgs state $A_S$. For a sufficiently
large ${\chi}^0_1$ pair annihilation cross section into two photons (and
$E_\gamma =  M_{{\chi}^0_1} \sim 130$~GeV), we need
\beq\label{eq:10}
M_{A_S} \sim 2M_{{\chi}^0_1}\sim 260~\text{GeV}\; ,
\eeq
which can be achieved by appropriate values of $-3  \kappa s A_\kappa$.
Moreover, the SU(2) doublet admixture of $A_S$ must be small: Otherwise
tree level diagrams similar to Fig.~\ref{fig:1} but with $A_S$ decaying
directly into $b\bar{b}$ (or into $Z$ plus a light CP-even Higgs boson),
lead to a too large pair annihilation cross section of ${\chi}^0_1$ such
that its relic density is below the WMAP bound. From the second of
Eqs.~\eqref{eq:7} the mixing of $A_S$ with the MSSM-like doublet is
small for
\beq\label{eq:11}
A_\lambda \approx 2\kappa s\; .
\eeq

Next we consider the dark matter particle ${\chi}^0_1$. It would have a
large singlino component for small $2\kappa s$. However, from
Eq.~\eqref{eq:6} and the first of Eqs.~\eqref{eq:7} one can derive
\beq\label{eq:12}
(2\kappa s)^2 \sim {\cal M}^2_{H\,33} + \frac{1}{3} M^2_{A_S}\; ;
\eeq
from $M_{A_S} \sim 260$~GeV and ${\cal M}^2_{H\,33}>0$ it follows that
$2\kappa s$ cannot be small. Hence, assuming $M_1 \lsim M_2/2$
(assuming universal gaugino masses at the GUT scale), it follows that
${\chi}^0_1$ has dominant bino and higgsino components. A priori large
higgsino components seem desirable, given the required coupling of
${\chi}^0_1$ to $A_S$ in Fig.~\ref{fig:1}: This coupling is induced by
the first term $\lambda \widehat S \widehat H_u\cdot \widehat H_d$ in
the superpotential \eqref{eq:1}, which leads to a Yukawa coupling
$\lambda A_S \tilde{H_u}^0 \tilde{H_d}^0$. Likewise, the coupling of
$A_S$ to the charginos $\chi^\pm$ originates from the higgsino
components $\tilde{H_u}^+$, $\tilde{H_d}^-$ of $\chi^\pm$ and the Yukawa
coupling $\lambda A_S \tilde{H_u}^+ \tilde{H_d}^-$.

However, too large higgsino components of ${\chi}^0_1$ imply again a too
small relic density; diagrams with charginos and neutralinos in the
t-channel (and $W^+ W^-$ or $Z Z$ in the final state), CP-even Higgs
bosons in the s-channel etc. would lead to a too large pair annihilation
cross section of ${\chi}^0_1$. Hence we end up with a dominantly
bino-like ${\chi}^0_1$, but with non-zero (non-negligible) higgsino
components. Its mass of 130~GeV has to follow from appropriate values of
$M_1$ and $\mu_{\text{eff}}$, with  $M_1<\mu_{\text{eff}}$.

Finally we require a SM-like CP-even Higgs boson $H_{SM}$ with a mass
$M_{H_{SM}}$ near 125~GeV. Although its existence is not confirmed at
present, it is interesting to investigate whether it could comply with
the above properties of the neutralino and CP-odd Higgs sector. It has
been known since a long time that the SM-like CP-even Higgs boson can be
heavier in the NMSSM compared to the MSSM due to the NMSSM-specific
coupling $\lambda S H_u H_d$ \cite{Ellis:1988er,Drees:1988fc,
Ellwanger:1993hn,Maniatis:2009re, Ellwanger:2009dp,Ellwanger:2006rm},
provided $\lambda$ is large and $\tan\beta$ is relatively small.
Subsequently we choose corresponding values of $\lambda$ and $\tan\beta$
such that $M_{H_{SM}}\sim 125$~GeV  \cite{Hall:2011aa,Arvanitaki:2011ck,
Ellwanger:2011aa, Gunion:2012zd, King:2012is,Kang:2012tn,Cao:2012fz,
Vasquez:2012hn,Ellwanger:2012ke, Lodone:2012kp,Jeong:2012ma}. We find
that the above properties in the neutralino and CP-odd Higgs
sector imply that $H_{SM}$ is the lightest CP-even Higgs state; the
singlet-like CP-even Higgs state has a mass $\gsim 200$~GeV. The
scenario with a lightest singlet-like Higgs state and a
next-to-lightest SM-like Higgs state (allowing for an enhanced branching
ratio into $\gamma \gamma$ \cite{Ellwanger:2011aa,King:2012is,
Cao:2012fz,Vasquez:2012hn}) seems difficult to realize. Herewith we have
sketched the interesting regions in the parameter space
in~\eqref{eq:5}. 

An open question remains whether the dark
matter particle can comply with the constraints from XENON100 on its
spin-independent detection cross section: ${\chi}^0_1$-nucleon
scattering is induced dominantly by $H_{SM}$-exchange in the t-channel
\cite{Underwood:2012ta}, and the $H_{SM}$-${\chi}^0_1$-${\chi}^0_1$
vertex is proportional to the product of the bino- and
higgsino-components of ${\chi}^0_1$ (from $g_1 \times$ bino $\times$
Higgs $\times$ higgsino terms in the Lagrangian, where $g_1$ is the
U(1)$_B$ gauge coupling). This issue will be studied below.

We have scanned the parameter space of the general NMSSM with help of
the NMSSMTools package \cite{Ellwanger:2004xm,Ellwanger:2005dv},
supplemented with suitably modified formulas for the cross sections for
${\chi}^0_1 {\chi}^0_1 \to \gamma \gamma$ (and for ${\chi}^0_1
{\chi}^0_1 \to Z \gamma$) from \cite{Bergstrom:1997fh, Ullio:1997ke}.
MicrOmegas \cite{Belanger:2005kh,Belanger:2006is,Belanger:2008sj} is
used for the calculation of the dark matter relic density and direct
detection cross sections. For the latter we have to specify the strange
quark content of the nucleons, i.e. the relevant sigma terms. We use the
most recent values from \cite{Thomas:2012tg} with, to be conservative, a
value for $\sigma_{\pi N}$ near the lower end of the $1\sigma$ error
bars: $\sigma_{\pi N} = 40$~MeV, $\sigma_0 = 39$~MeV.

\vspace*{3mm}

\noindent
For the soft SUSY breaking terms we made the following choices:
\begin{itemize}
\item Squark masses of 1.5~TeV, except for the left-handed squarks of
the 3rd generation (1~TeV) and the right-handed top squark (300~GeV).
The latter values are motivated by universal soft scalar masses at the
GUT scale \cite{Ellwanger:2012ke}, and alleviate LHC constraints from
direct SUSY searches due to the more complicated squark and gluino decay
cascades involving the light stops.
\item  Trilinear soft susy breaking terms $A_t = A_b = -1.1$~TeV.
\item  Slepton masses in the $140-500$~GeV range such that the SUSY
contributions to the anomalous muon magnetic moment are sufficiently
large (inspite of low values of $\tan\beta$), while slepton
exchange in the t-channel of the  pair annihilation cross section of
${\chi}^0_1$ does not imply a too small relic density. 
\item  Whereas we vary $M_1$ in the $140-160$~GeV range (see below),
$M_2$ and the gluino mass $M_3$ are kept fixed at $M_2=300$~GeV,
$M_3=800$~GeV for simplicity.
\item Finally we use 173.1~GeV for the top quark pole mass.
\end{itemize}

\vspace*{3mm}
\noindent
Subsequently we impose the following phenomenological constraints:
\begin{itemize}
\item $M_{{\chi}^0_1} = 129 - 131$~GeV and $\langle \sigma
v\rangle({\chi}^0_1 {\chi}^0_1 \to \gamma \gamma) > 10^{-27}\
\text{cm}^3\ \text{s}^{-1}$ in order to obtain a photon line in
agreement with the excess found in \cite{Weniger:2012tx,Tempel:2012ey}. 

\item Upper bounds on annihilation cross sections into $W^+ W^-$, $Z Z$,
$b\bar{b}$ and $\tau \bar{\tau}$ channels from the Fermi~LAT
collaboration \cite{Ackermann:2012qk,Ackermann:2011wa} (see also
\cite{Hooper:2011ti,1205.6811}), as well as bounds from PAMELA on the
antiproton flux \cite{Adriani:2010rc}. For the determination of these
cross sections/fluxes we use micrOMEGAs2.4 \cite{Belanger:2010gh}.

\item A relic density complying with the WMAP bound $\Omega h^2 =
0.1120\pm 0.011$ \cite{Komatsu:2010fb} (with $2 \sigma$ error bars).
\item A SM-like Higgs boson with $M_{H_{SM}} = 124 - 127$~GeV, as it was
confirmed recently by the ATLAS and CMS collaborations
\cite{Atlas-Higgs, CMS-Higgs}.

\item Constraints from B-physics as implemented in NMSSMTools (which
have actually no impact for the low values of $\tan\beta$ considered
here).
\item A sufficiently large SUSY contribution $\Delta a_\mu$ to the muon
anomalous magnetic moment as implemented in NMSSMTools.
\item Constraints from the absence of Landau singularities of the
running Yukawa couplings below the GUT scale, and the absence of
unphysical global minima of the Higgs potential.
\end{itemize}

\begin{figure}[hb!]
\vspace*{-5mm}
\begin{center}
\includegraphics[scale=0.51,angle=-90]{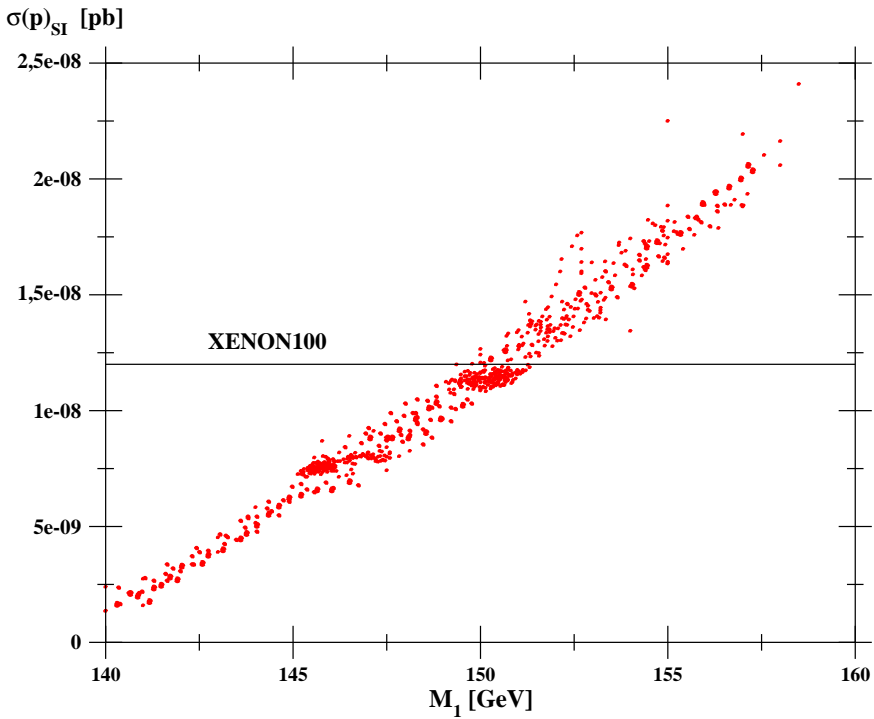}
\end{center}
\vspace*{-5mm}
\caption{$\sigma(p)_{SI}$ as function of $M_1$ for a sample of points
satisfying all other phenomenological constraints as indicated in the
text. The horizontal line indicates the present bound from XENON100 for
a 130~GeV dark matter particle, which holds for all points shown.}
\label{fig:2}
\end{figure}

We have found corresponding points in the NMSSM parameter space, both
below and slightly above the present XENON100 bound on the
spin-independent dark matter -- proton cross section of $\sigma(p)_{SI}
\lsim 1.2 \times 10^{-8}$~pb for $M_{{\chi}^0_1} \sim
130$~GeV~\cite{Aprile:2011hi}. In Fig.~\ref{fig:2} we show
$\sigma(p)_{SI}$ as function of $M_1$ for a sample of such points, where
we varied the parameters in \eqref{eq:5} in the range
$\lambda = 0.6-0.615$, $\kappa= 0.326-0.329$, $A_\lambda= 240-400$~GeV,
$A_\kappa= -130 - (-60)$~GeV, $\mu_{\text{eff}}= 230-445$~GeV,
$\tan\beta = 1.68-1.82$.

\begin{table}[ht!]
\begin{center}
\begin{tabular}{|c|c|c|} \hline
\multicolumn{2}{|c|} {Parameters}  \\\hline
$\lambda$               & 0.61      \\\hline
$\kappa$                & 0.328   \\\hline
$A_\lambda$             &  267    \\\hline
$A_\kappa$              &  -114.1 \\\hline
$\tan\beta$             & 1.8    \\\hline
$\mu_{\text{eff}}$      & 269    \\\hline
$M_1$                   & 150    \\\hline
left-h. slepton masses  & 150    \\\hline
right-h. slepton masses & 160    \\\hline
$A_e=A_\mu=A_\tau$ & 500    \\\hline
\hline
\multicolumn{2}{|c|} {Sparticle masses}    \\\hline
$m_{\tilde{g}}$         &  971       \\\hline
$\left<m_{\tilde{q}}\right>$ &   1530   \\\hline
$m_{\tilde{t}_1}$     &   204   \\\hline
$m_{\tilde{t}_2}$     &   1034   \\\hline
$m_{\tilde{b}_1}$     &   1005   \\\hline
$m_{\tilde{\mu}_L}$   &   154  \\\hline
$M_{\chi^0_1}$            &   129.6 \\\hline
$M_{\chi^0_2}$            &   217 \\\hline
$M_{\chi^0_3}$            &   287 \\\hline
$M_{\chi^0_4}$            &   309 \\\hline
$M_{\chi^0_5}$            &   376 \\\hline
$M_{\chi^\pm_1}$      &    210 \\\hline
$M_{\chi^\pm_2}$      &    370 \\\hline
\end{tabular}
\hspace*{10mm}
\begin{tabular}{|c|c|c|} \hline
\multicolumn{2}{|c|} {Higgs masses}    \\\hline
$M_{H_1} (=M_{H_{SM}})$  & 124.3   \\\hline
$M_{H_2}$                & 256     \\\hline
$M_{H_3}$                & 519     \\\hline
$M_{A_1} (=M_{A_{S}})$   & 258.9   \\\hline
$R^{bb}_{A_{S}}$         & $3\times 10^{-3}$  \\\hline
$M_{A_2}$                & 515   \\\hline
$M_{H^\pm}$              & 511     \\\hline
\hline
\multicolumn{2}{|c|} {Components of $\chi^0_1$}       \\\hline
$N_{11}^2$               & 0.826 \\\hline
$N_{12}^2$               & 0.026 \\\hline
$N_{13}^2$               & 0.077 \\\hline
$N_{14}^2$               & 0.065 \\\hline
$N_{15}^2$               & 0.009 \\\hline
\hline
\multicolumn{2}{|c|} {Observables}       \\\hline
$\Omega h^2$            &   0.11  \\\hline
$\sigma(p)_{SI}\ [10^{-8}$~pb] &   1.21    \\\hline
$\langle \sigma v\rangle({\chi}^0_1 {\chi}^0_1\to \gamma \gamma)\ 
[10^{-27}\text{cm}^3\ \text{s}^{-1}]$ & 1.1  \\\hline
$\langle \sigma v\rangle({\chi}^0_1 {\chi}^0_1\to Z \gamma)\ 
[10^{-27}\text{cm}^3\ \text{s}^{-1}]$ & 0.8  \\\hline
$\langle \sigma v\rangle({\chi}^0_1 {\chi}^0_1\to WW)\ 
[10^{-27}\text{cm}^3\ \text{s}^{-1}]$ & 3.46  \\\hline
$\langle \sigma v\rangle({\chi}^0_1 {\chi}^0_1\to Z Z)\ 
[10^{-27}\text{cm}^3\ \text{s}^{-1}]$ & 0.26  \\\hline
$\langle \sigma v\rangle({\chi}^0_1 {\chi}^0_1\to b \bar{b})\ 
[10^{-27}\text{cm}^3\ \text{s}^{-1}]$ & 0.60  \\\hline
$\langle \sigma v\rangle({\chi}^0_1 {\chi}^0_1\to \tau \bar{\tau})\ 
[10^{-27}\text{cm}^3\ \text{s}^{-1}]$ & 0.09  \\\hline
$\Delta a_\mu\ [10^{-10}]$  & $6.5\pm 3.0$ \\\hline
\end{tabular}
\vspace*{10mm}
\caption{Properties of a sample point with $M_1=150$~GeV. Dimensionful
parameters are given in GeV. $R^{bb}_{A_{S}}$
denotes the coupling of ${A_{S}}$ to $b$-quarks normalized to the one of
a SM-like Higgs boson. The components of $\chi^0_1$ are defined in
Eq.~\eqref{eq:9}. The value of $\Delta a_\mu$ includes theoretical
error bars.}
\end{center}
\end{table}

In Table~1 we show the details (parameters, masses and relevant
observables) for a sample point with $M_1 = 150$~GeV. The couplings
of $H_1 \equiv H_{SM}$ to quarks, leptons, electroweak gauge bosons,
gluons and photons are SM-like within $\sim 5\%$. $R^{bb}_{A_{S}}$
denotes the coupling of ${A_{S}}$ to $b$-quarks normalized to the one of
a SM-like Higgs boson; its small value underlines its singlet-like
nature.

The following remarks are in order: The larger $M_1$, the smaller one
has to choose $\mu_{\text{eff}}$ in order to maintain
$M_{{\chi}^0_1} \sim 130$~GeV. It follows that, for larger $M_1$, the
higgsino component of ${\chi}^0_1$ increases leading to a larger
${\chi}^0_1 {\chi}^0_1$ annihilation cross section. Hence, for too large
$M_1$, the relic density falls below the WMAP bound.

\begin{figure}[ht!]
\begin{center}
\includegraphics[scale=0.6,angle=-90]{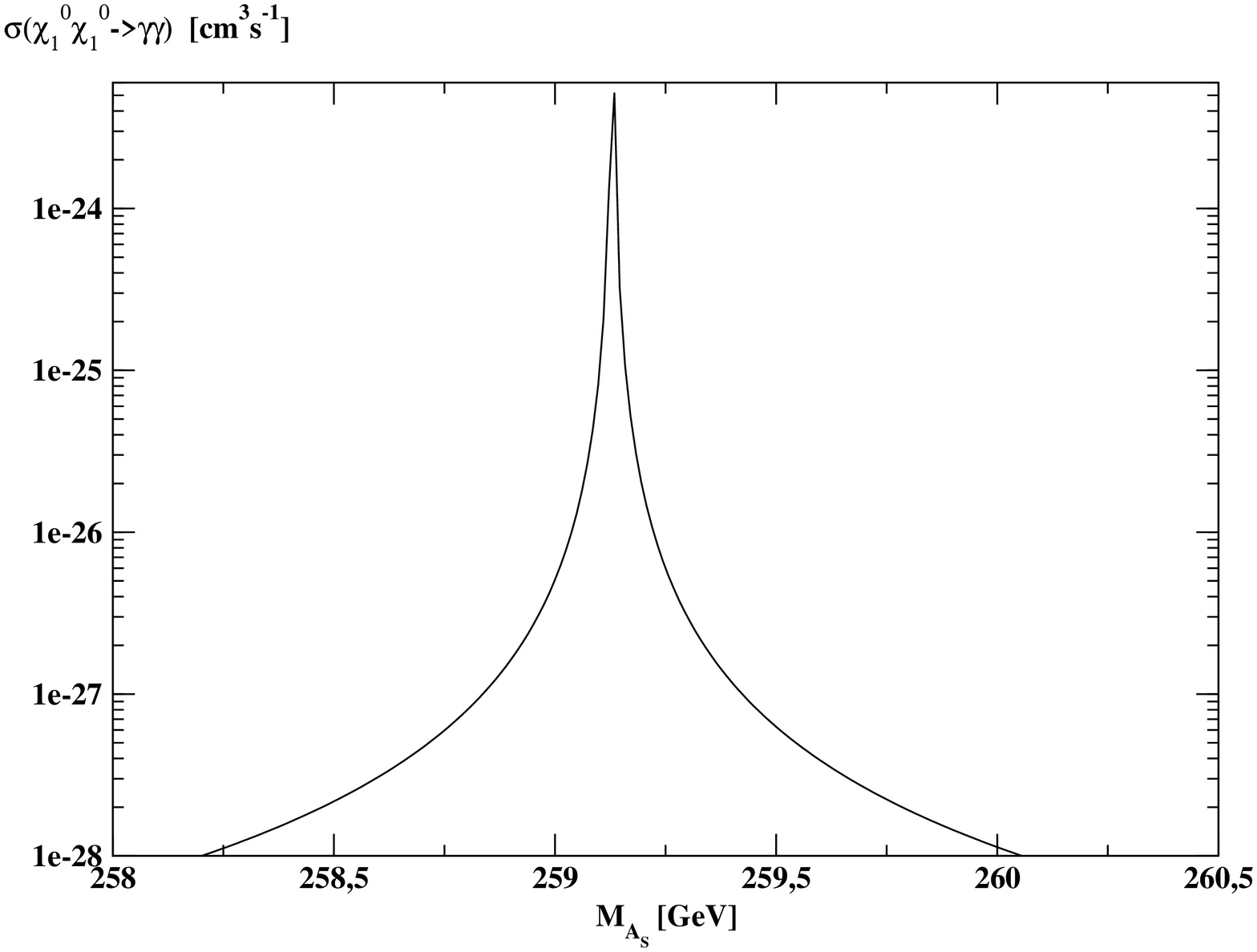}
\end{center}
\vspace*{-5mm}
\caption{$\sigma({\chi}^0_1 {\chi}^0_1\to \gamma \gamma)$ as function of
$M_{A_S}$ for the point listed in Table~1.}
\label{fig:3}
\end{figure}

On the other hand, for smaller $M_1$ one has to choose larger values for
$\mu_{\text{eff}}$ implying a smaller higgsino component of
${\chi}^0_1$, which explains the decrease of $\sigma(p)_{SI}$ in
Fig.~\ref{fig:2}. However,
simultaneously the coupling of $A_S$ to ${\chi}^0_1$ decreases as well.
As a consequence, the mass $M_{A_S}$ of $A_S$ has to be closer and
closer to the pole $2 M_{{\chi}^0_1}$ in order to obtain
$\sigma({\chi}^0_1 {\chi}^0_1 \to \gamma \gamma) > 10^{-27}\
\text{cm}^3\ \text{s}^{-1}$. In order to clarify the required tuning, we
show $\sigma({\chi}^0_1 {\chi}^0_1\to \gamma \gamma)$ in
Fig.~\ref{fig:3} as function of $M_{A_S}$ for the point listed in
Table~1. (Note
that the finite width $\Gamma(A_S) \sim 1.6$~KeV, dominated by $A_S \to
\gamma\gamma$, is not visible in Fig.~\ref{fig:3}. Due to the small
couplings of $A_S$ to quarks and leptons, the contributions of $A_S$ in
the s-channel to the annihilation cross sections into $b\bar{b}$ and
$\tau \bar{\tau}$ final states are well below the Fermi~LAT bounds. The
annihilation cross sections into $W^+ W^-$ and $Z Z$ originate from the
second CP-even Higgs boson in the s-channel. The antiproton flux has a
maximum of $\sim 2.47\times 10^{-4}\;(\text{GeV}\,\text{m}^2\,
\text{s}\,\text{sr})^{-1}$ for an energy of $\sim 2 \; \text{GeV}$,
which is well below the PAMELA bound.)

We see that $\sigma({\chi}^0_1 {\chi}^0_1\to \gamma \gamma)$ is larger
than $10^{-27}\ \text{cm}^3\ \text{s}^{-1}$ only within a $\sim 0.7$~GeV
wide window of $M_{A_S}$. This required tuning becomes worse for lower
values of $M_1$, and would be the price to pay for a possibly stronger
constraint on $\sigma(p)_{SI}$ from XENON100 in the future. (On the
other hand, modifications of the present best estimates for the
Higgs-nucleon coupling and/or the local dark matter density could
alleviate the present constraints from XENON100.)

Finally we should add that diagrams similar to Fig.~\ref{fig:1}, but
with one photon replaced by a $Z$ boson, contribute to
$\sigma({\chi}^0_1 {\chi}^0_1\to Z \gamma)$ leading to an additional
photon line with, for $M_{{\chi}^0_1}\sim 130$~GeV, $E_\gamma \sim
114$~GeV. For the present scenario we find $\sigma({\chi}^0_1
{\chi}^0_1\to Z \gamma) \sim 75\% \times \sigma({\chi}^0_1 {\chi}^0_1\to
\gamma \gamma)$. Such an additional line would be compatible with the
structure observed in \cite{1206.1616}. In any case, additional
lines---also from $\sigma({\chi}^0_1 {\chi}^0_1\to H \gamma)$ or
interpreting the 130~GeV line as due to ${\chi}^0_1 {\chi}^0_1\to Z
\gamma$---can be interesting checks of such scenarios in the future
\cite{Rajaraman:2012db}.

\section{Conclusions}

In this paper we have shown that the simplest version of the NMSSM (with
a scale invariant superpotential) could explain a 130~GeV photon line
from dark matter annihilation with $\sigma({\chi}^0_1 {\chi}^0_1 \to
\gamma \gamma) > 10^{-27}\ \text{cm}^3\ \text{s}^{-1}$ and,
simultaneously, a 125~GeV SM-like Higgs boson. No additional fields or
couplings need to be introduced. All constraints from WMAP on the relic
density, from XENON100 on the direct detection cross section, from
colliders and from precision observables can be satisfied.

However, the mass $M_{A_S}$ of the singlet-like CP-odd Higgs scalar
$A_S$ has to satisfy accidentially $M_{A_S} \approx 2 M_{{\chi}^0_1}
\sim 260$~GeV to a precision $\lsim 1$~GeV. This ``fine-tuning'' would
become worse if bounds on the direct detection cross section become
stronger (but could be relaxed otherwise).

Unfortunately a direct verification of this scenario at colliders
through searches for a 130~GeV photon line seems hopeless: Due to the
singlet-like nature of $A_S$, production cross sections for this state
(as well as decay widths of sparticles or other Higgs bosons into this
state) are too small. Only the mass of 130~GeV of the LSP ${\chi}^0_1$
should fit the data, once searches for supersymmetry turn out to be
successful. Of course, first of all the present hints for a 130~GeV
photon line \cite{Bringmann:2012vr,Weniger:2012tx,Tempel:2012ey,
{1206.1616}} need to be confirmed.

\section*{Acknowledgements}
We would like to thank Y. Mambrini for discussions.
U.~E. acknowledges partial support from the French ANR LFV-CPV-LHC, ANR
STR-COSMO and the European Union FP7 ITN INVISIBLES (Marie Curie
Actions,~PITN-GA-2011-289442). P.~M. acknowledges support from the
Greek State Scholarship Foundation.


\end{document}